\documentclass[aip,jcp,reprint]{revtex4-2}

\usepackage{amsmath}
\usepackage{graphicx}
\usepackage{dcolumn}
\usepackage{bm}
\usepackage[utf8]{inputenc}
\usepackage[T1]{fontenc}
\usepackage{mathptmx}
\usepackage{etoolbox}
\usepackage[normalem]{ulem}
\usepackage{mhchem}
\usepackage{xcolor}
\usepackage{hyperref}
\hypersetup{hidelinks}

\makeatletter
\def\@email#1#2{%
	\endgroup
	\patchcmd{\titleblock@produce}
	{\frontmatter@RRAPformat}
	{\frontmatter@RRAPformat{\produce@RRAP{*#1\href{mailto:#2}{#2}}}\frontmatter@RRAPformat}
	{}{}
}%
\makeatother

\begin{document}
	\raggedbottom
	
	\title{Unifying Time-Independent and Time-Dependent
		Franck--Condon Pictures in Quantum Mechanochemistry
		via the Linear Coupling Approximation}
	\thanks{The following article has been submitted to
		\textit{The Journal of Chemical Physics}. The following article has been submitted to
			\textit{The Journal of Chemical Physics}.
			Copyright 2026 Vladimir V. Rybkin. This article is distributed under
			a Creative Commons Attribution (CC BY) License.}
	
	\author{Vladimir V. Rybkin}
	\affiliation{HQS Quantum Simulations GmbH,
		Rintheimer Stra\ss{}e~23,
		D-76131 Karlsruhe, Germany}
	\email{your.email@example.com}
	
	\begin{abstract}
	Quantum chemistry has achieved impressive results in modeling and
	interpreting mechanochemical transformations at the molecular level.
	An essential question, the time profile of force application, has
	however attracted less attention. Here, the Franck--Condon theory
	of quantum mechanochemistry, previously formulated for the
	sudden-force limit, is generalized to arbitrary time-dependent
	force profiles using the linear-coupling approximation. Solving the
	driven harmonic oscillator yields a closed-form, time-dependent
	effective Huang--Rhys factor: a natural measure of vibrational excitations upon force application. It is governed by the Fourier transform of
	the force derivative and recovers the sudden- and fixed-force limits. At the conceptual level, this linear
	coupling picture links mechanochemistry to vibronic spectroscopy and Marcus electron
	theory. For a shock-front force ramp, vibrational excitation is
	shown to be suppressed in frequency, only
	low-frequency modes being excited under realistic laser-shock-peening
	conditions.
	\end{abstract}
	
	\maketitle
	
	%
	%
	
	\section{Introduction}
	
	Mechanical force has emerged as an important tool for activating
	chemical reactions~\cite{Beyer_chemrev,Pagola_review}.
	It can be applied using different techniques, including
	ultrasound~\cite{cravotto2013ultrasound},
	ball mills~\cite{ballmilling2025primer},
	single-molecule manipulation~\cite{neuman2008single} via
	atomic-force microscopy~\cite{rief1997reversible} or optical
	tweezers~\cite{kellermayer1997folding},
	shock waves~\cite{dlott2011new} (including laser shock peening,
	LSP~\cite{fairand1979laser,peyre1995laser}), and other methods.
	Understanding these phenomena at the level of molecular quantum
	chemistry has progressed significantly in recent
	years~\cite{stauch2016advances,alonso2024wandering,RibasArino2012}.
	
	The seminal work of Bell~\cite{bell1978models} established the
	linear dependence of the reaction barrier on the applied force
	and thus explained mechanochemical activation in a simple and
	compelling way. This idea was elaborated by including nonlinear
	terms in the force dependence of the barrier~\cite{Konda2011}.
	An alternative approach models force application by constrained
	geometry optimization (COGEF), leaving some intramolecular
	forces unrelaxed to mimic the external force~\cite{Beyer2000}.
	
	Another generalization of Bell's approach applied a linear
	force term to the entire potential energy surface (PES),
	yielding the force-transformed, or force-modified, PES
	(FM-PES)~\cite{RibasArino2009}. This idea has proven
	particularly fruitful because it allows the smooth application
	of well-established PES-exploration machinery to the FM-PES,
	including geometry optimization, transition-state theory (TST),
	and molecular dynamics (MD), including steered MD. In addition,
	several ingenious analysis techniques and concepts have arisen,
	including judgement of energy distribution (JEDI)
	analysis~\cite{Stauch2014}, Newton-trajectory
	analysis~\cite{Quapp2016}, and pulling
	corridors~\cite{Quapp2017}. Importantly, the equivalence of
	explicit force inclusion (EFEI) and COGEF was
	established~\cite{Beyer2000}; these approaches are related by a
	Legendre transformation, with force and internal coordinate
	being conjugate variables~\cite{RibasArino2009}.
	
	The FM-PES and EFEI approaches are agnostic to the nature of
	the force and usually treat it as constant in time. Several
	contributions have addressed the first issue by explicitly
	including the force in the electronic
	Hamiltonian~\cite{Wolinski2009} and by computing hydrostatic
	stress through the electronic
	Hamiltonian~\cite{Stauch2020,Weiss2024}. These approaches bring
	mechanochemical models closer, at least in principle, to the
	physical conditions of real experiments. Regarding time
	dependence, a notable application considered a stepwise linear
	force ramp corresponding to the load rate estimated for
	ultrasound~\cite{Stauch2017}.
	
	The basis for the systematic study of varying force was laid
	down by Smal{\o}~\textit{et al.}, who identified two limiting
	cases: fixed force and sudden force
	application~\cite{Smalo2014}. The former is adiabatically slow,
	whereas the latter assumes the immediate application of the
	force at its full strength, in a manner analogous to a vertical
	electronic transition. This idea was developed further by the
	author of this contribution by recognizing that sudden
	(``vertical'') external force application corresponds to the
	Condon principle, in which the transition takes place not
	between different electronic states, but rather between the PES
	and the FM-PES~\cite{Rybkin2017}. Consequently, in addition to
	barrier lowering, further activation occurs through vibrational
	excitation on the FM-PES, which can be described using
	Franck--Condon (FC) factors. This framework explains the
	different fragmentation patterns of butane in the fixed- and
	sudden-force regimes, as well as the faster rates in the latter
	regime.
	
	Realistic modeling of different experiments requires different
	force time profiles. Single-molecule force spectroscopy is
	almost certainly adiabatic, or close to the fixed-force limit,
	because the force ramps are many orders of magnitude slower
	than vibrational relaxation times. However, it is unclear
	whether the fixed-force regime corresponds to any experimental
	reality under current technological conditions. Importantly,
	the issue of unrealistically fast loading rates in classical
	biomolecular simulations of single-molecule force spectroscopy
	has been identified and discussed~\cite{Sheridan2019}. In
	quantum chemistry, however, the force-application protocol is
	often implicit: sudden force application can occur in molecular
	MD applications~\cite{Ong2019}, whereas in \textit{ab~initio}
	MD simulations of materials and liquids under periodic boundary
	conditions, fixed-force models based on COGEF-like constraints,
	such as fixing terminal atoms or changing the simulation cell,
	are commonly used~\cite{Dopieralski2013,Nepal2021}. In studies
	using static approaches such as TST, fixed-force application is
	typically assumed~\cite{Smalo2013,Pill2019}.
	
	In this work, I attempt to emphasize the role of the
	force-loading rate by generalizing the FC theory of quantum
	mechanochemistry to time-dependent external-force application.
	In doing so, the FC theory is simplified using the linear
	coupling approximation (LC). This gives several notable results
	linking mechanochemistry to spectroscopy and electron-transfer
	theory, even in the fixed-force regime. The final result is an
	expression for the time-dependent Huang--Rhys factors, which
	define the FC factors in the LC approximation, for a general
	force-ramp profile. This expression is used to estimate the
	amount of vibrational excitation under typical experimental
	conditions.
	
	%
	%
	
	\section{Time-Independent Theory}
	
	\subsection{Harmonic Force-Modified PES as a
		Linear-Coupling Model}
	
\begin{figure}[htbp]
		\centering
		\includegraphics[width=\linewidth]{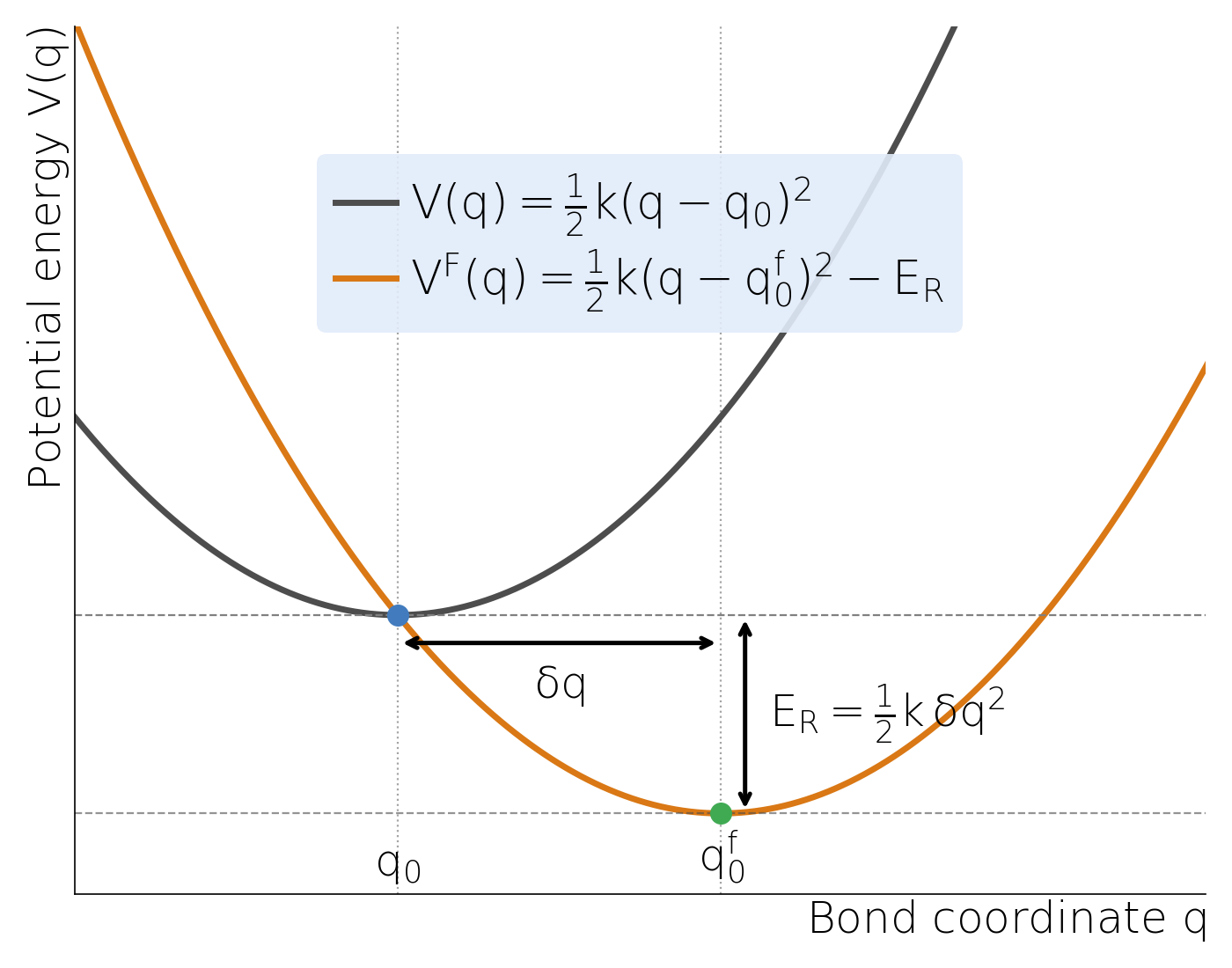}
		\caption{Force-modified PES as a linear-coupling model:
			gray,~$V(q)$, harmonic PES; orange,~$V(q)^f$, force-modified
			PES (FM-PES); $\delta q$, coordinate displacement; $q_0$ and
			$q_0^f$, minima on the PES and FM-PES, respectively; $E_R$,
			reorganization energy; $k$, harmonic force constant.}
		\label{fig:parabolas}
	\end{figure}

	For a one-dimensional case, such as a diatomic molecule, the
	force-modified potential energy surface (FM-PES) in the
	harmonic approximation can be written as
	\begin{equation}
		V(q)
		= \frac{1}{2}k(q-q_0)^2 - f(q-q_0),
		\label{FM_1d}
	\end{equation}
	where $q$ is the coordinate, $q_0$ is its equilibrium value in
	the absence of an external force, $k$ is the harmonic force
	constant, and $f$ is the force magnitude.
	Equation~\eqref{FM_1d} differs from a standard harmonic PES,
	$V(q)=\frac{1}{2}k(q-q_0)^2$, by the linear force-dependent
	term. This term allows the application of a linear-coupling
	model known from vibronic spectroscopy without further
	approximation~\cite{frank1975electron}, as shown in
	Fig.~\ref{fig:parabolas}.
	
	Application of an external force shifts the PES minimum to a
	new position:
	\begin{equation}
		q_0^f = q_0 + \frac{f}{k} = q_0 + \delta q.
		\label{minimum}
	\end{equation}
	
	If the force application is instantaneous, or
	sudden~\cite{Rybkin2017}, the Condon
	approximation~\cite{condon1926theory,frank1975electron} applies
	and leads to a nonthermal population of vibrational energy
	levels on the FM-PES. The probabilities of vibrational
	excitation following sudden force application are given by
	Franck--Condon factors (FCFs). These are overlaps between the
	vibrational wave functions on the original PES,~$\chi_n$, and
	those on the FM-PES,~$\chi_m^f$.
	
	%
	%
	
	Evaluating FCFs in the linear-coupling model is
	straightforward because the force constant remains the same for
	the PES and the FM-PES. Therefore, the vibrational wave
	functions associated with the PES and FM-PES are harmonic
	oscillator eigenfunctions related by a shift of origin. For
	simplicity, we consider the zero-temperature case, in which
	only the vibrational ground state is populated before force
	application. We introduce the dimensionless displacement
	parameter
	\begin{equation}
		\Delta = \sqrt{\frac{\mu\omega}{\hbar}}\,\delta q,
		\label{eq:Delta}
	\end{equation}
	where $\mu$ is the reduced mass, $\omega$ is the angular
	vibrational frequency, and $\hbar$ is the reduced Planck
	constant. The probability of a transition from $\chi_0$ to
	$\chi_m^f$, the Franck--Condon factor, is then
	\begin{equation}
		\begin{aligned}
			F_{m0}
			&= \left|\langle\chi_m^f\mid\chi_0\rangle\right|^2 \\
			&= \frac{(\Delta^2/2)^m}{m!}e^{-\Delta^2/2}
			= \frac{S^m}{m!}e^{-S}, \\
			S &= \Delta^2/2,
		\end{aligned}
		\label{eq:Poisson}
	\end{equation}
	where $S$ is the Huang--Rhys
	factor~\cite{Huang1950TheoryOL}. One can show that the
	Huang--Rhys factor equals the molecular reorganization energy
	following sudden force application, expressed in units of the
	vibrational quantum:
	\begin{equation}
		E_R = S\hbar\omega.
		\label{eq:reorg}
	\end{equation}
	
	Since the vibrational levels are equidistant in the harmonic
	approximation, $S$ equals the average number of vibrational
	quanta in the state under the external force, or the level of
	vibrational excitation. Substituting Eq.~\eqref{eq:Delta}
	into Eq.~\eqref{eq:reorg}, we obtain
	\begin{equation}
		\begin{aligned}
			E_R
			&= \frac{1}{2}\mu\omega^2\delta q^2 \\
			&= \frac{1}{2}\frac{f^2}{k}.
		\end{aligned}
		\label{eq:reorg_force}
	\end{equation}
	
	The reorganization energy,
	$E_R = \frac{1}{2}\mu\omega^2\delta q^2 = \frac{1}{2}k\delta q^2$,
	is the same as the strain energy, which is the harmonic
	increase in energy at finite displacement, analogous to the
	energy required to stretch a spring.
	Equation~\eqref{eq:reorg} establishes the dependence of the
	reorganization, or strain, energy on the applied force, which
	is quadratic.
	
	The force dependence of the Huang--Rhys factor is
	\begin{equation}
		S = \frac{1}{2}\frac{\mu\omega}{\hbar}\delta q^2
		= \frac{1}{2}\frac{\mu\omega}{\hbar}\frac{f^2}{k^2}
		= \frac{f^2}{2\hbar^2\mu\omega^3}.
		\label{eq:static_S}
	\end{equation}
	
	It is worth noting that $E_R$ in the two-parabola picture in
	Fig.~\ref{fig:parabolas} is conceptually analogous to the
	Marcus reorganization energy in electron-transfer
	processes~\cite{Marcus1956}.
	
	%
	%
	
	\subsection{Kinetics on the Force-Modified PES}
	
	Although a harmonic potential does not describe bond breaking,
	we assume that it provides a good approximation for vibrational
	motion, while the dissociation limit on the PES is still
	present, as is standard in this type of model. According to
	the simple Bell theory~\cite{bell1978models}, the dissociation
	barrier, or activation energy $E_A^f$, depends linearly on the
	applied force:
	\begin{equation}
		E_A^f = E_A - f(q-q_0).
	\end{equation}
	
	Thus, the larger the pulling force, the lower the dissociation
	barrier. This observation provides the simplest explanation of
	mechanical activation. In the sudden-force regime, additional
	activation originates from vibrational excitation, as
	quantified by the Huang--Rhys factor
	[Eq.~\eqref{eq:reorg}]. Within a simple Fridman--Macheret
	model, this excess energy, or strain/reorganization energy, is
	subtracted from the activation
	energy~\cite{1981UsFiN.134..185R,macheret1993chemphys}:
	\begin{equation}
		E_A^f(\text{sudden})
		= E_A^f - E_R
		= E_A^f - S\hbar\omega.
		\label{eq:Fridman}
	\end{equation}
	
	Combining Eqs.~\eqref{eq:Fridman} and~\eqref{eq:reorg_force},
	we arrive at the activation-energy dependence on the external
	force in the sudden regime within the linear-coupling
	approximation:
	\begin{equation}
		E_A^f(\text{sudden})
		= E_A - f(q-q_0) - \frac{1}{2}\frac{f^2}{k}.
		\label{activation}
	\end{equation}
	
	The second term is present only in the sudden regime. It is
	more sensitive to the force magnitude than the barrier lowering
	in the linear Bell picture, although the more realistic model
	of Konda~\textit{et al.} assumes a quadratic dependence of the
	activation energy on the external force~\cite{Konda2011}.
	
	%
	%
	
	\subsection{Generalization to Polyatomic Molecules}
	
	For a polyatomic molecule, the FM-PES in the basis of normal
	modes~$\mathbf{q}$ reads
	\begin{equation}
		V(\mathbf{q})
		= \frac{1}{2}(\mathbf{q}-\mathbf{q}_0)^{T}
		\mathbf{k}(\mathbf{q}-\mathbf{q}_0)
		- \mathbf{f}^{T}(\mathbf{q}-\mathbf{q}_0),
	\end{equation}
	where $\mathbf{k}$ is the diagonal force-constant matrix and
	$\mathbf{f}$ is the force vector in the normal-coordinate
	frame.
	
	The force-shifted minimum of the PES becomes
	\begin{equation}
		\mathbf{q}_0^F = \mathbf{q}_0 + \mathbf{k}^{-1}\mathbf{f},
	\end{equation}
	where $\mathbf{k}^{-1}$ is also known as the compliance
	matrix.
	
	The reorganization energy is a sum of contributions from the
	individual vibrational modes:
	\begin{equation}
		E_R
		= \frac{1}{2}\sum_i m_i\omega_i^2\delta q_i^2
		= \sum_i S_i\hbar\omega_i
		= \sum_i E_R^i.
		\label{reorg_poly}
	\end{equation}
	
	This equation is equivalent to the central equation of the
JEDI analysis written in
	normal coordinates. As established previously, the strain
	energy in JEDI equals the linear-coupling reorganization
	energy~$E_R$. This highlights the quantum nature of the
	molecule, in which $E_R$ has a discrete spectrum for each
	particular external force~$\mathbf{f}$, with the values
	of~$E_i$ forming a Franck--Condon profile.
	
	The equivalence of strain energy and reorganization energy
	allows the apparatus of linear-coupling Franck--Condon theory
	to be applied not only to sudden-force cases, but also to
	fixed-force cases and static strain analysis, because the
	strain energy can be expressed in terms of Huang--Rhys factors.
	
	The magnitude of the Huang--Rhys factor~$S_i$ for each
	vibration depends on the force magnitude and its direction,
	through the projection of the force onto the normal mode. It
	may happen that the external force~$\mathbf{f}$ has no
	component along the normal mode corresponding to the reaction
	coordinate, or that it compresses the bond, thereby increasing
	the barrier~$E_A$ rather than decreasing it. Even in this
	case, sudden force application can still accelerate the
	reaction by creating vibrational excitation, that is, by
	heating the molecule. Although the harmonic model considered
	here cannot \textit{per se} explain intramolecular vibrational
	energy redistribution, such redistribution will still occur
	from excited modes with large force projections into
	reaction-relevant modes that are not directly excited by the
	initial force application.
	
	%
	%
	
	\section{Time-Dependent Theory}
	
	\subsection{Time-Dependent Hamiltonian and Basis}
	
	If the external force is time-dependent, the Hamiltonian takes the form
	\begin{equation}
		\hat{H}(t)
		=
		\underbrace{
			\frac{\hat{p}^2}{2\mu}
			+ \frac{1}{2}k(q-q_0)^2
		}_{\hat{H}_0}
		- f(t)(q-q_0),
		\label{eq:H}
	\end{equation}
	where $\hat{H}_0$ is the unperturbed harmonic-oscillator
	Hamiltonian. Equation~\eqref{eq:H} defines the well-known
	problem of the driven harmonic oscillator.
	
	Although its solution is generally
	known~\cite{cohen-tannoudji_qm_vol1}, we present it here using
	notation tailored to mechanochemical applications. The force
	ramps from $f(-\infty)=0$ to $f(+\infty)=f_{\max}$. In the
	sudden-force regime, the force is switched on immediately as a
	step function, whereas in the fixed-force regime it increases
	adiabatically slowly. These are limiting cases; here we solve
	the problem for a general force profile~$f(t)$.
	
	Following the previous derivation [Eq.~\eqref{minimum}], the
	instantaneous FM-PES minimum is
	\begin{equation}
		q_0^{f(t)} = q_0 + \frac{f(t)}{k}.
	\end{equation}
	
	The final FM-PES at $f=f_{\max}$ has eigenstates~$\chi_m^f$
	with total equilibrium displacement
	\begin{equation}
		\delta q = \frac{f_{\max}}{k},
		\qquad
		\Delta = \sqrt{\frac{\mu\omega}{\hbar}}\,\delta q.
	\end{equation}
	
	We are interested in the vibrational distribution over the
	states~$\chi_m^f$. For a driven harmonic system, this
	distribution is Poissonian~\cite{lapierre2018getting}:
	\begin{equation}
		P_m = e^{-S_{\mathrm{eff}}}\frac{S_{\mathrm{eff}}^m}{m!},
		\qquad
		\langle m\rangle = S_{\mathrm{eff}},
	\end{equation}
	fully parameterized by~$S_{\mathrm{eff}}$. The static result
	gives the sudden-limit value for~$f_{\max}$.
	
	It is convenient to write the Hamiltonian in second
	quantization. We introduce the bosonic ladder operators
	\begin{equation}
		\hat{a}
		= \sqrt{\frac{\mu\omega}{2\hbar}}
		\left[(q-q_0)+\frac{i\hat{p}}{\mu\omega}\right],
		\quad
		q-q_0
		= \sqrt{\frac{\hbar}{2\mu\omega}}\,(\hat{a}+\hat{a}^{\dagger}),
	\end{equation}
	which give the Hamiltonian
	\begin{equation}
		\hat{H}(t)
		= \hbar\omega\!\left(\hat{a}^{\dagger}\hat{a}+\frac{1}{2}\right)
		- f(t)\sqrt{\frac{\hbar}{2\mu\omega}}
		\,(\hat{a}+\hat{a}^{\dagger}).
		\label{sq_Hamiltonian}
	\end{equation}
	
	\subsection{Effective Huang--Rhys Factor}
	
	The FM-PES Hamiltonian has the same force constant~$k$ and
	frequency~$\omega$ as the original, but is shifted by~$\delta
	q$. Its ladder operator is $\hat{b}=\hat{a}-\Delta$.
	
	The FM-PES ground state $|\chi_0^f\rangle$ is the vacuum state
	of~$\hat{b}$:
	\begin{equation}
		\hat{b}|\chi_0^f\rangle = 0
		\;\Longrightarrow\;
		\hat{a}|\chi_0^f\rangle = \Delta|\chi_0^f\rangle,
	\end{equation}
	so $|\chi_0^f\rangle$ is a coherent state of~$\hat{a}$ with
	amplitude~$\Delta$:
	\begin{equation}
		|\chi_0^f\rangle
		= |\Delta\rangle
		= \hat{D}(\Delta)|\chi_0\rangle,
		\qquad
		\hat{D}(\Delta)
		= \exp\!\left(\Delta\hat{a}^{\dagger}-\Delta^{*}\hat{a}\right),
	\end{equation}
	where $|\chi_0\rangle$ is the ground vibrational state of the
	original, force-free PES and $\hat{D}(\Delta)$ is the
	displacement operator. All FM-PES Fock states $|\chi_m^f\rangle$
	are created from this coherent ground state by repeated
	application of $\hat{b}^{\dagger}=\hat{a}^{\dagger}-\Delta$:
	\begin{equation}
		|\chi_m^f\rangle
		= \frac{(\hat{b}^{\dagger})^m}{\sqrt{m!}}|\chi_0^f\rangle
		= \frac{(\hat{a}^{\dagger}-\Delta)^m}{\sqrt{m!}}
		\hat{D}(\Delta)|\chi_0\rangle.
	\end{equation}
	
	Since $\hat{H}(t)$ is linear in $\hat{a}$ and
	$\hat{a}^{\dagger}$, the driven state remains a coherent
	state $|\Delta(t)\rangle$ of~$\hat{a}$ at all times:
	\begin{equation}
		\hat{a}|\Delta(t)\rangle = \Delta(t)|\Delta(t)\rangle,
		\label{eq:Delta_interact}
	\end{equation}
	where $\Delta(t)$ is the Schr\"odinger-picture coherent
	amplitude. Moving to the interaction picture removes the free
	precession:
	\begin{equation}
		\widetilde{\Delta}(t) = \Delta(t)\,e^{i\omega t}.
	\end{equation}
	
	For the FM-PES, the ground-state amplitude in the interaction
	picture is
	\begin{equation}
		\widetilde{\Delta}_{\mathrm{FM}}(t) = \Delta\,e^{i\omega t}.
		\label{eq:DeltaFM}
	\end{equation}
	
	The Huang--Rhys factor is related to the displacement
	parameter in the corresponding displacement operator
	[Eq.~\eqref{eq:Poisson}]~\cite{Eickmann2026} and measures
	vibrational excitation on the final FM-PES relative to its
	ground state~$|\chi_0^f\rangle$. Here it is the squared
	distance between the driven amplitude and the FM-PES
	ground-state amplitude, both in the interaction picture:
	\begin{equation}
		S_{\mathrm{eff}}
		= \lim_{t\to\infty}
		\left|\widetilde{\Delta}(t)
		- \widetilde{\Delta}_{\mathrm{FM}}(t)\right|^2
		= \lim_{t\to\infty}
		\left|\widetilde{\Delta}(t)
		- \Delta\,e^{i\omega t}\right|^2.
		\label{eq:Seff_def}
	\end{equation}
	
	Solving the Heisenberg equation of motion, as detailed in
	Appendix~\ref{app:derivation}, gives the central result:
	\begin{equation}
		S_{\mathrm{eff}}(\omega)
		= \frac{\left|\widetilde{\dot{f}}(\omega)\right|^2}
		{2\mu\hbar\omega^3},
		\qquad
		\widetilde{\dot{f}}(\omega)
		= \int_{-\infty}^{\infty}\dot{f}(t)\,e^{i\omega t}\,dt.
		\label{eq:Seff_final}
	\end{equation}
	
	The excitation is governed by the power spectrum of the force
	derivative, filtered by~$\omega^{-3}$. Only the time variation
	of the force produces vibrational excitation on the FM-PES.
	
	\subsection{Limiting and Intermediate Cases}
	
	In the sudden-force case, $f(t)$ is a Heaviside step function:
	\begin{equation}
		f(t) = f_{\max}\Theta(t),
		\qquad
		\Theta(t) =
		\begin{cases}
			0, & t < 0, \\
			1, & t > 0.
		\end{cases}
	\end{equation}
	
	Its time derivative is
	\begin{equation}
		\dot{f}(t) = f_{\max}\delta(t)
		\;\Longrightarrow\;
		\widetilde{\dot{f}}(\omega) = f_{\max}.
	\end{equation}
	
	Therefore,
	\begin{equation}
		S_{\mathrm{eff}}^{\mathrm{sudden}}(\omega)
		= \frac{f_{\max}^2}{2\mu\hbar\omega^3}
		= S,
	\end{equation}
	recovering the static Franck--Condon result of
	Eq.~\eqref{eq:static_S}.
	
	In the adiabatic limit, the ramp is infinitely slow, so
	$\dot{f}(t)\to 0$ at all finite frequencies:
	\begin{equation}
		\widetilde{\dot{f}}(\omega)\to 0
		\;\Longrightarrow\;
		S_{\mathrm{eff}}^{\mathrm{adiabatic}}(\omega) = 0.
	\end{equation}
	The molecule follows the FM-PES adiabatically, with no
	vibrational excitation.
	
\begin{figure}[htbp]
		\centering
		\includegraphics[width=\linewidth]{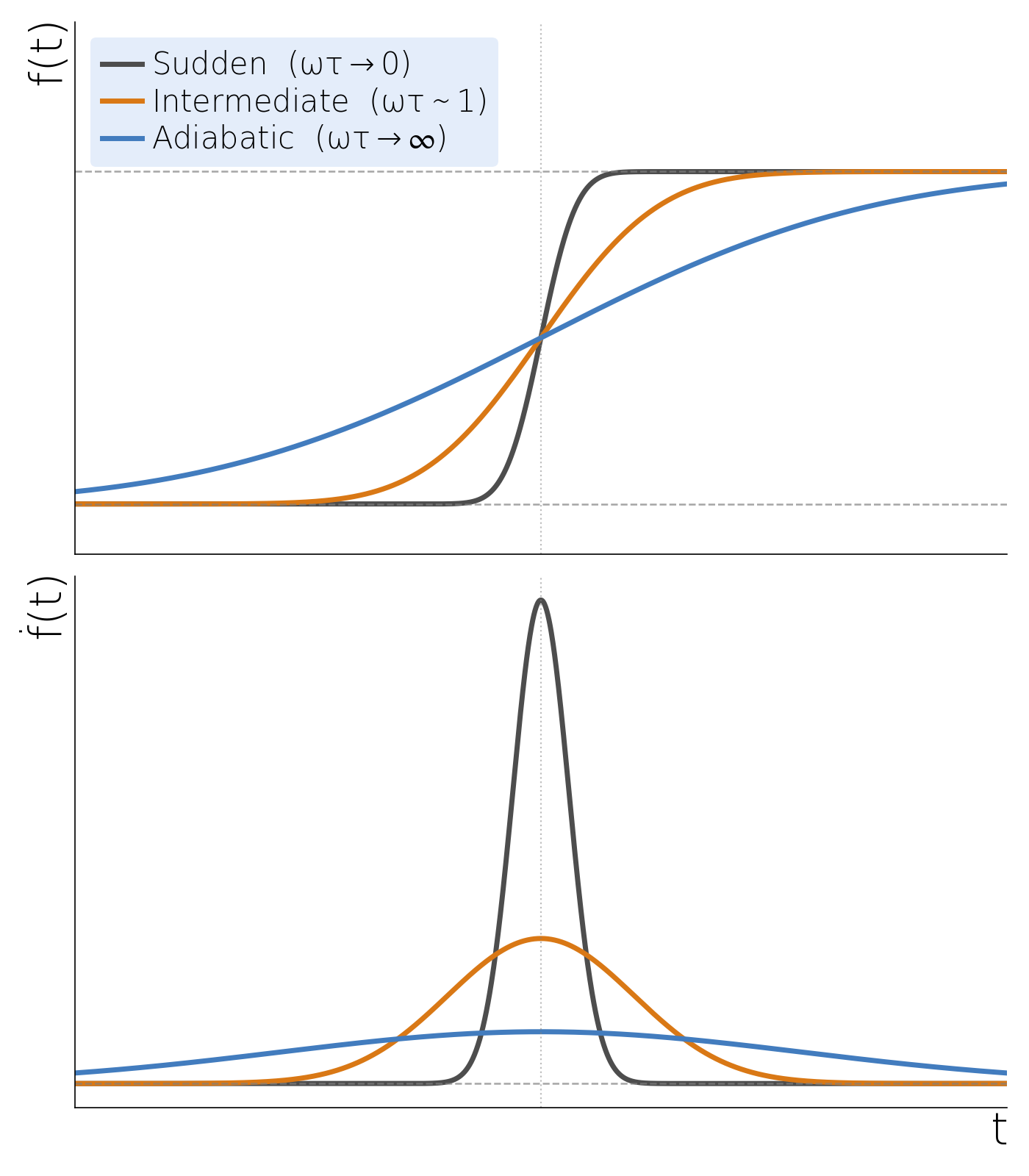}
		\caption{Force-ramp profiles (upper panel), shown as error
			functions, and their time derivatives (lower panel) for
			different regimes: gray, sudden; blue, adiabatic or fixed;
			orange, intermediate.}
		\label{fig:force_ramps}
	\end{figure}

	For the physically motivated shock-front profile with rise
	time~$\tau$ (see Fig.~\ref{fig:force_ramps} and
	Table~\ref{tab:regimes}):
	\begin{equation}
		f(t)
		= \frac{f_{\max}}{2}
		\left[1+\operatorname{erf}\!\left(
		\frac{t}{\sqrt{2}\,\tau}\right)\right].
	\end{equation}
	
	Its time derivative is a Gaussian,
	\begin{equation}
		\dot{f}(t)
		= \frac{f_{\max}}{\sqrt{2\pi}\,\tau}
		\exp\!\left(-\frac{t^2}{2\tau^2}\right),
	\end{equation}
	with Fourier transform
	\begin{equation}
		\widetilde{\dot{f}}(\omega)
		= f_{\max}\exp\!\left(-\frac{\omega^2\tau^2}{2}\right).
	\end{equation}
	
	Substitution into Eq.~\eqref{eq:Seff_final} gives
	\begin{equation}
		S_{\mathrm{eff}}(\omega)
		= S\exp\!\left(-\omega^2\tau^2\right).
		\label{eq:Seff_erf}
	\end{equation}
	
	%
	%
	
	%
	%
	
	\begin{table}[htbp]
		\centering
		\caption{Limiting behavior of the effective Huang--Rhys
			factor for the error-function force ramp.}
		\label{tab:regimes}
		\begin{ruledtabular}
			\begin{tabular}{lcc}
				Limit & Condition & $S_{\mathrm{eff}}(\omega)$ \\
				\hline
				Sudden   & $\omega\tau\to 0$        & $S$ \\
				General  & $\omega\tau$ finite      & $S\,e^{-\omega^2\tau^2}$ \\
				Adiabatic& $\omega\tau\to\infty$    & $0$ \\
			\end{tabular}
		\end{ruledtabular}
	\end{table}
	
	The Gaussian filter $e^{-\omega^2\tau^2}$ is the central
	result. High-frequency modes, for which $\omega\tau\gg1$, are
	exponentially suppressed, whereas low-frequency modes, for
	which $\omega\tau\ll1$, retain the sudden-limit excitation to a
	large degree. For a laser-shock front with
	$\tau\sim50$--$100\,\mathrm{fs}$, this suppression becomes
	significant already around
	$\tilde{\nu}\sim300$--$500\,\mathrm{cm}^{-1}$, confirming that
	only low-frequency normal modes are efficiently excited by the
	mechanical shock.
	
	\section{Applications}
	\subsection{Numerical Estimates with Molecular Parameters}
	
	Here, we compute effective Huang--Rhys factors for molecular
	vibrations in intermediate force-ramp regimes using parameters
	derived from LSP experiments. The parameters for the
	error-function force-ramp profile, $f_{\max}$ and~$\tau$, are
	estimated in Appendix~\ref{app:estimation}. Because the kinetics
	are exponential, the logarithmic scale used in the right-hand
	panels gives an indication of the vibrational contributions to
	the activation energy in Eq.~\eqref{eq:Fridman}.
	
\begin{figure*}[htbp]
		\centering
		\includegraphics[width=\textwidth]{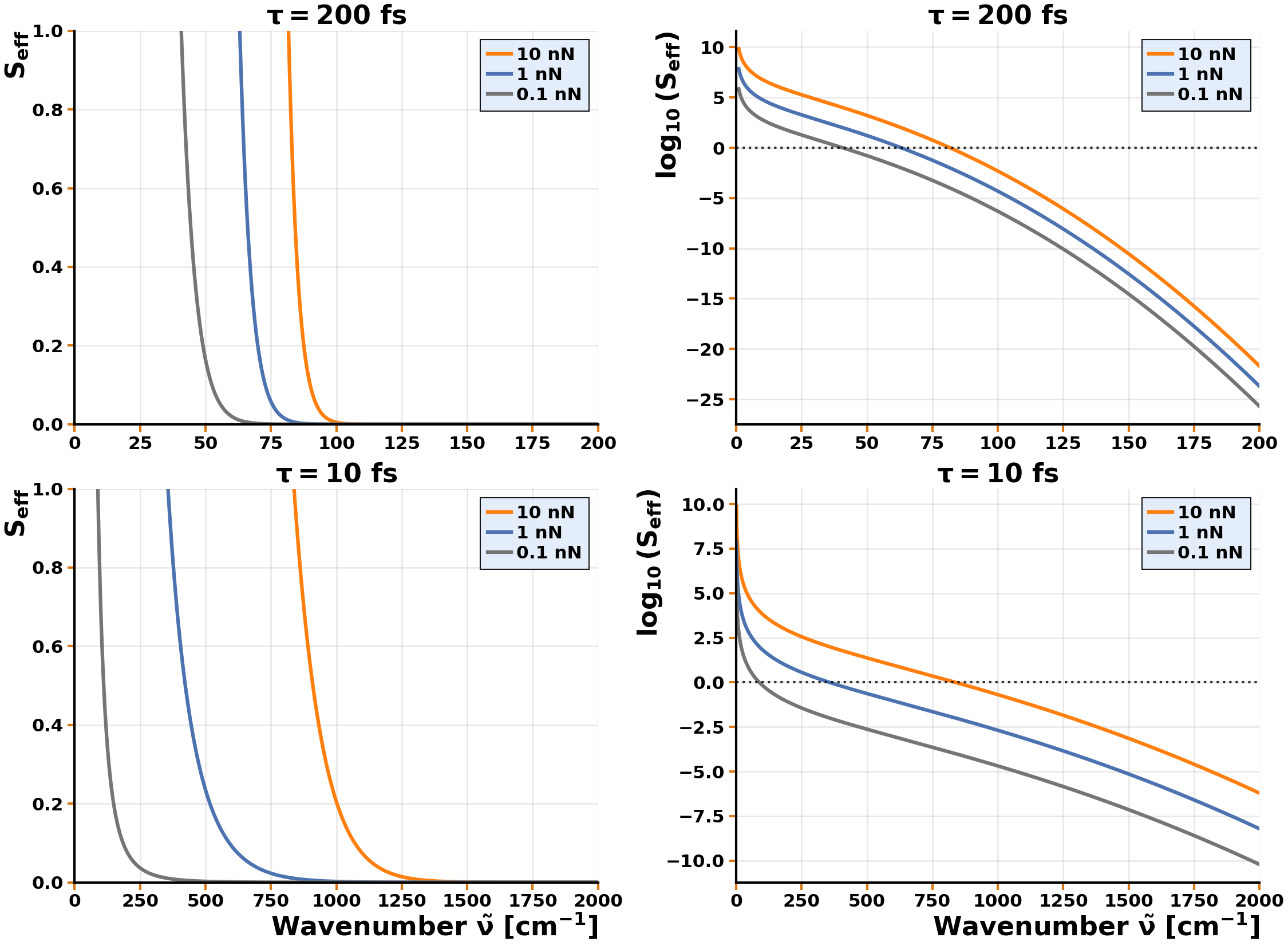}
		\caption{Effective Huang--Rhys factors as a function of
			molecular vibrational frequency for different force values
			and ramp times.}
		\label{fig:Seff}
	\end{figure*}

	Results for several parameter sets are shown in
	Fig.~\ref{fig:Seff}. At moderate force time ramp values lower-frequency vibrations acquire
	significant vibrational excitation, namely bending, torsional and acoustic
	modes in the solid state. This excess vibrational energy can be
	transferred to bond-stretching vibrations through mechanisms of
	intermolecular vibrational redistribution; estimating such
	effects requires quantum-dynamics simulations and is system
	dependent.
	
	%
	%
	
	%
	%
	
\subsection{Stretching \textit{n}-Butane}
\begin{figure*}[htbp]
		\centering
		\begin{minipage}[t]{0.32\textwidth}
			\centering
			\includegraphics[width=\linewidth]{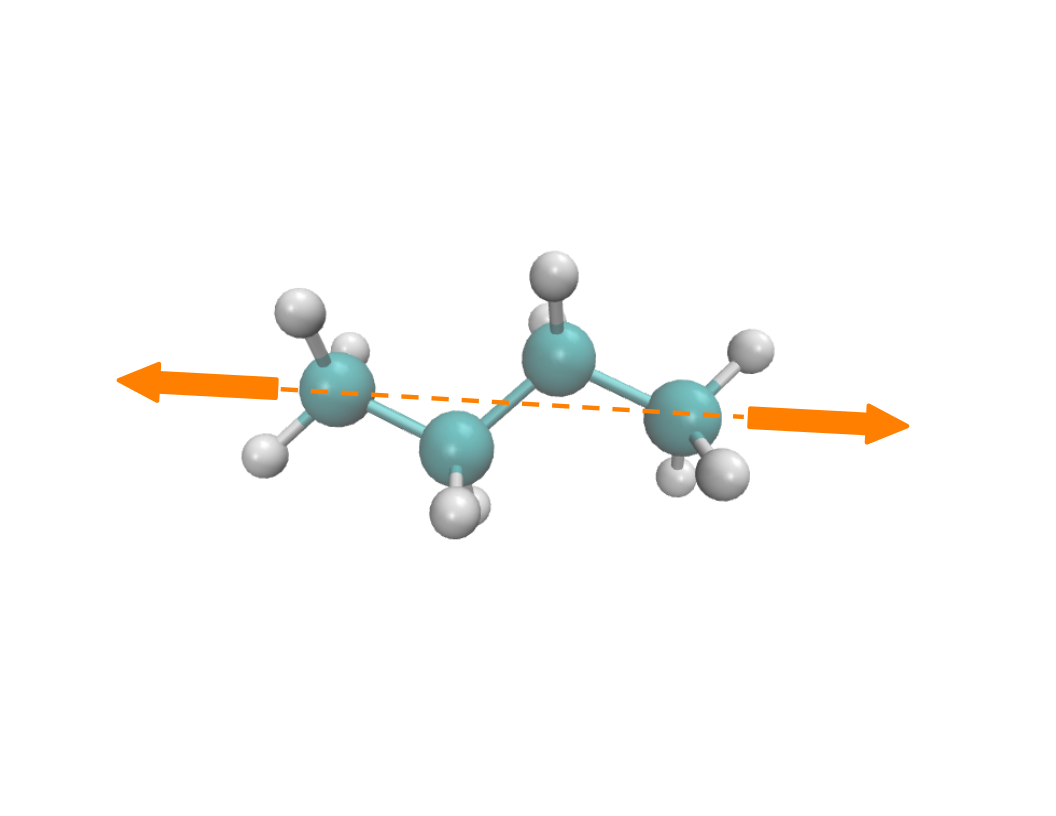}\\
			(a)
		\end{minipage}\hfill
		\begin{minipage}[t]{0.33\textwidth}
			\centering
			\includegraphics[width=\linewidth]{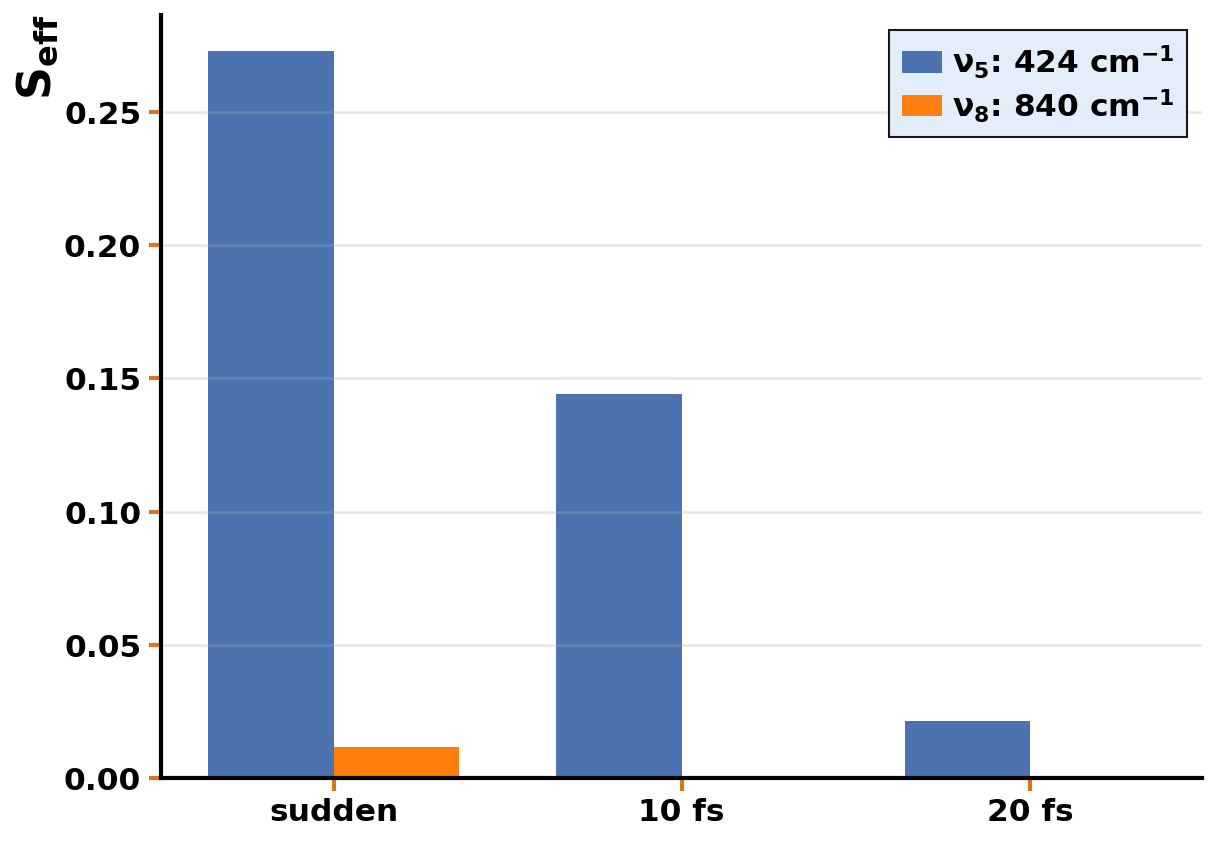}\\
			(b)
		\end{minipage}\hfill
		\begin{minipage}[t]{0.33\textwidth}
			\centering
			\includegraphics[width=\linewidth]{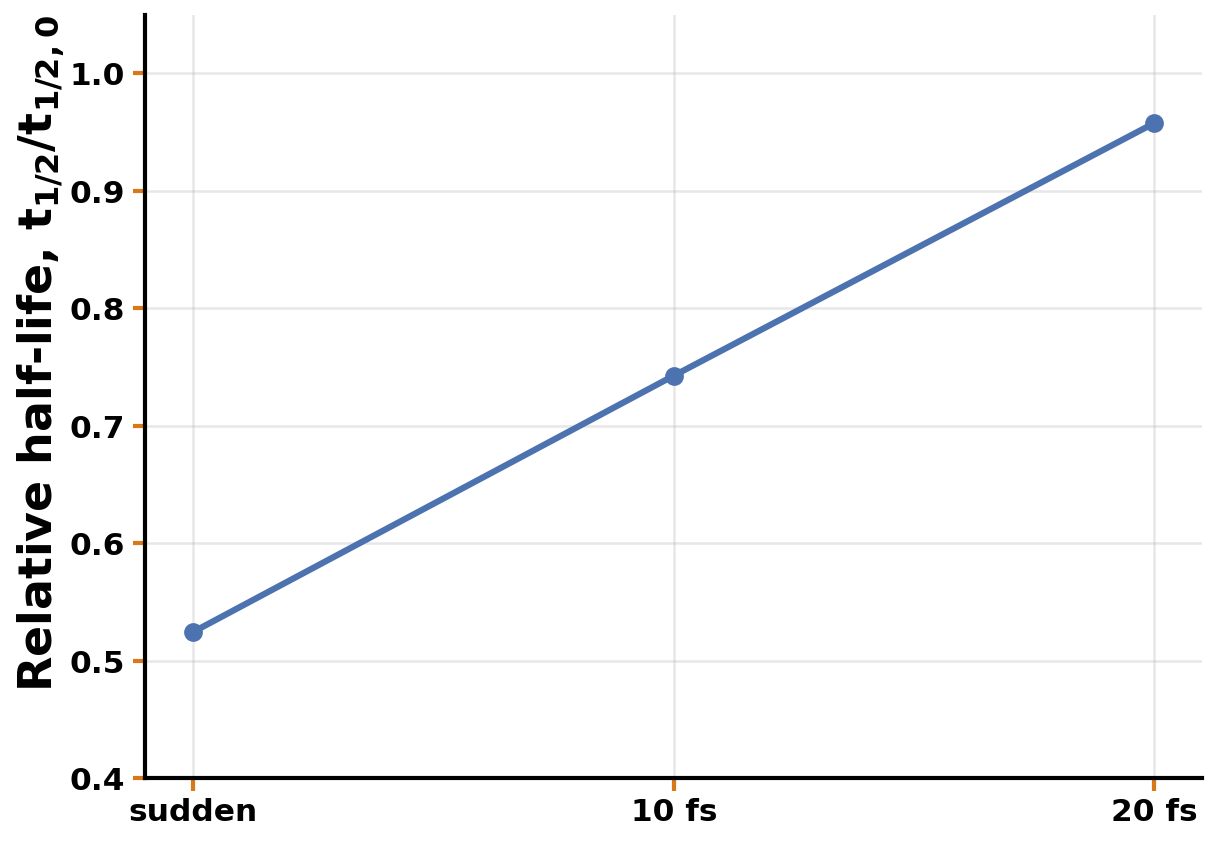}\\
			(c)
		\end{minipage}
		\caption{Application to \textit{n}-butane: (a) stretching of the
			\textit{n}-butane molecule by external force of 2.95 nN
			; (b) effective Huang--Rhys factors of
			the selected normal modes; (c) relative half-life for
			different force-ramp times.}
		\label{fig:butane}
	\end{figure*}

Next, we present an application of the theory to the mechanical stretching of \textit{n}-butane, which was considered in previous works \cite{Smalo2014, Rybkin2017}. For consistency, we used the same protocol and quantum-chemical approach: DFT at the B3LYP\cite{Becke1993, Lee1988}/6-311G**\cite{Krishnan1980} level. The geometry was first optimized without constraints and in the absence of applied forces, followed by a standard Wilson harmonic analysis\cite{Wilson1955} using PySCF\cite{Sun2018}. We then applied an external stretching force along the line connecting the two terminal carbon atoms (see Figure \ref{fig:butane}(a)). The force magnitude was $2.95~\mathrm{nN}$, corresponding to $0.75$ of the critical value of $3.9~\mathrm{nN}$ determined elsewhere \cite{Smalo2014}. This was the maximum force value used for the static analysis in the first study of Franck--Condon theory \cite{Rybkin2017}.

Time-dependent Huang--Rhys factors were subsequently calculated for ramp times of $10$ and $20~\mathrm{fs}$ using Equation \eqref{eq:Seff_erf}. The sudden-limit values were evaluated using Equation \eqref{eq:static_S}, with the force components determined by projecting the external force onto the normal modes.

The results are shown in Figure \ref{fig:butane}. As expected, significant Huang--Rhys factors are obtained for only two modes: the symmetric chain-bending (deformation) mode, with $\nu_5 = 424~\mathrm{cm}^{-1}$, and the symmetric stretching mode, with $\nu_8 = 840~\mathrm{cm}^{-1}$. For nonzero $\tau$, however, the contribution of the latter becomes negligible. We observe a drastic decrease in $S$ with increasing ramp time.

To illustrate the effect of these vibrational excitations on the reaction kinetics, relative first-order reaction half-lives were calculated using Equation \eqref{eq:Fridman} for the activation energy and the Arrhenius equation, assuming that both the pre-exponential factor and the static energy barrier are independent of the ramp time. Owing to the exponential dependence in the Arrhenius equation, even a modest additional activation can lead to considerable acceleration, as shown in panel~(c) of Figure \ref{fig:butane}.
	
	%
	%
	
	%
	%
	
	\section{Conclusions and Outlook}
	
	Introducing the linear-coupling approximation into the
	Franck--Condon theory of quantum mechanochemistry, previously
	formulated for sudden force application, allows this
	vibronic-like picture to be generalized to adiabatic
	(fixed-force) and intermediate force-application regimes.
	The main conceptual conclusion of this work is the dependence of
	 mechanochemical activation efficiency not only on the magnitude and
	 direction of the external force, but also on the rate and time profile
	 of its application. These may result in additional vibrational excitations 
	 accelerating mechanochemical reactions: such an effect is in principle measurable
	 provided force-ramp profile $f(t)$, direction and magnitude are 
	 controllable. Figure~\ref{fig:butane} illustrates these
	effects for \textit{n}-butane.
	
	The amount of additional vibrational excitation, or heating,
	arising from external mechanical force is characterized by the
	effective time-dependent Huang--Rhys factor,
	$S_{\mathrm{eff}}$, which depends on the time-ramp profile of
	the force~$f(t)$ and its time derivative~$\dot{f}(t)$,
	sometimes called the yank. In the limiting cases of fixed and
	sudden force action, $S_{\mathrm{eff}}$ reaches its extreme
	values: zero and its maximum value, respectively.
	
	For fixed force, the linear-coupling approximation for the
	harmonic PES is exactly equivalent to the force-modified PES
	model with a linear force term, which is the most commonly used
	approach in mechanochemical theory and applications. This makes
	it possible to establish terminological connections between
	spectroscopy and mechanochemistry. In particular, the strain
	energy is equivalent to the reorganization energy of a vibronic
	transition, and also to that associated with charge transfer.
	The contribution of each vibrational mode is given by the
	corresponding Huang--Rhys factor.
	
	Estimation of $S_{\mathrm{eff}}$ for typical LSP experiments and computing those for stretching of \textit{n}-butane
	show that direct excitation of valence-bond-stretching
	vibrations is unlikely under currently accessible ramp times
	and force magnitudes. Such a regime may remain futuristic;
	therefore, care must be taken when relating theoretical
	calculations that implicitly adopt sudden force application to
	experiments. On the other hand, bending, torsional and acoustic
	vibrations, as well as slow modes in
	biomolecules~\cite{Sheridan2019}, are likely to undergo
	non-negligible excitation and can affect the kinetics through
	intermolecular vibrational
	redistribution~\cite{tokmakoff1993chemical}. Other types of
	experiments, such as femtosecond
	shocks~\cite{Evans1996}, different ramp functions, and advances
	in LSP techniques may reveal a different picture. More
	realistic theoretical models incorporating anharmonicity and
	quantum dynamics will likely show a more nuanced behavior.
	
	At this point a special note must be made regarding
	mechanochemical reactions in solution \cite{James2012}. Due to high collision rates
	vibrational energy dissipates rapidly in this environment, which may
	 mitigate vibrational heating discussed here. On the other hand,
	 energy transfer from excited solvent molecules to the substrate is
	 also possible opening interesting opportunities for theorizing and 
	 practice.
	
	I believe that this simple model will be useful for qualitative
	estimation of finite-time force-application effects in
	mechanochemistry, whereas the terminological and conceptual
	conclusions may stimulate further understanding of the field.
	
	\begin{acknowledgments}
	This work has received funding from the German Federal Ministry of
	Research, Technology and Space (BMFTR) within the PhoQuant project
	(Grant No.~13N16103).
	\end{acknowledgments}

	\section*{Author Declarations}

	\subsection*{Conflict of Interest}
	The author has no conflicts to disclose.

	\subsection*{Author Contributions}
	\textbf{Vladimir V. Rybkin}: Conceptualization, Formal analysis,
	Investigation, Methodology, Visualization, Writing -- original draft,
	Writing -- review \& editing.

	\section*{Data Availability}
	Data sharing is not applicable to this article as no new data were
	created or analyzed in this study.

	%
	%
	
	\appendix
	
	\section{Derivation of the Effective Huang--Rhys Factor Using
		the Heisenberg Equation of Motion}
	\label{app:derivation}
	
	The Heisenberg equation for the annihilation operator,
	$d\hat{a}/dt=(i/\hbar)[\hat{H}(t),\hat{a}]$, takes the form
	\begin{equation}
		\frac{d\hat{a}}{dt}
		= -i\omega\hat{a}
		+ \frac{if(t)}{\sqrt{2\mu\hbar\omega}},
		\label{eq:heisenberg}
	\end{equation}
	after using the standard commutator relations
	$[\hat{a}^{\dagger}\hat{a},\hat{a}]=-\hat{a}$ and
	$[\hat{a}^{\dagger},\hat{a}]=-1$.
	
	The coherent amplitude is the expectation value of the
	annihilation operator, so its time derivative is
	\begin{equation}
		\dot{\Delta}(t)
		= -i\omega\Delta(t)
		+ \frac{if(t)}{\sqrt{2\mu\hbar\omega}}.
		\label{eq:Delta_dot}
	\end{equation}
	
	Differentiating the interaction-picture amplitude and
	substituting Eq.~\eqref{eq:Delta_dot}, we obtain
	\begin{equation}
		\begin{aligned}
			\dot{\widetilde{\Delta}}(t)
			&= \dot{\Delta}(t)e^{i\omega t}
			+ i\omega\Delta(t)e^{i\omega t} \\
			&= \left(-i\omega\Delta(t)
			+ \frac{if(t)}{\sqrt{2\mu\hbar\omega}}\right)e^{i\omega t}
			+ i\omega\Delta(t)e^{i\omega t}.
		\end{aligned}
	\end{equation}
	
	The $\pm i\omega\Delta(t)$ terms cancel exactly:
	\begin{equation}
		\dot{\widetilde{\Delta}}(t)
		= \frac{if(t)}{\sqrt{2\mu\hbar\omega}}\,e^{i\omega t}.
		\label{eq:Delta_tilde_dot}
	\end{equation}
	
	Integrating from $-\infty$ to $t$ with the initial condition
	$\widetilde{\Delta}(-\infty)=0$ gives
	\begin{equation}
		\widetilde{\Delta}(t)
		= \frac{i}{\sqrt{2\mu\hbar\omega}}
		\int_{-\infty}^{t}f(t')e^{i\omega t'}\,dt'.
		\label{eq:Delta_tilde}
	\end{equation}
	
	Integration by parts yields
	\begin{equation}
		\int_{-\infty}^{t}f(t')e^{i\omega t'}\,dt'
		= \left.\frac{f(t')e^{i\omega t'}}{i\omega}\right|_{-\infty}^{t}
		- \frac{1}{i\omega}
		\int_{-\infty}^{t}\dot{f}(t')e^{i\omega t'}\,dt'.
		\label{eq:ibp}
	\end{equation}
	
	The lower limit vanishes because $f(-\infty)=0$. Using
	$i/\sqrt{2\mu\hbar\omega}\cdot1/(i\omega)
	=1/\sqrt{2\mu\hbar\omega^3}$, we arrive at
	\begin{equation}
		\widetilde{\Delta}(t)
		= \frac{f(t)e^{i\omega t}}{\omega\sqrt{2\mu\hbar\omega}}
		+ \frac{1}{\sqrt{2\mu\hbar\omega^3}}
		\int_{-\infty}^{t}\dot{f}(t')e^{i\omega t'}\,dt'.
		\label{eq:Delta_ibp}
	\end{equation}
	
	Substituting Eq.~\eqref{eq:Delta_ibp} into
	Eq.~\eqref{eq:Seff_def}:
	\begin{equation}
		\begin{aligned}
			\widetilde{\Delta}(t) - \Delta e^{i\omega t}
			&= \frac{(f(t)-f_{\max})e^{i\omega t}}
			{\omega\sqrt{2\mu\hbar\omega}} \\
			&\quad+ \frac{1}{\sqrt{2\mu\hbar\omega^3}}
			\int_{-\infty}^{t}\dot{f}(t')e^{i\omega t'}\,dt'.
		\end{aligned}
		\label{eq:diff}
	\end{equation}
	
	As $t\to\infty$, $f(t)\to f_{\max}$, so the first term
	vanishes. The remaining integral converges to the full Fourier
	transform of~$\dot{f}$, giving our central result:
	\begin{equation}
		S_{\mathrm{eff}}(\omega)
		= \frac{\left|\widetilde{\dot{f}}(\omega)\right|^2}
		{2\mu\hbar\omega^3},
		\qquad
		\widetilde{\dot{f}}(\omega)
		= \int_{-\infty}^{\infty}\dot{f}(t)e^{i\omega t}\,dt.
		\label{eq:Seff_final_app}
	\end{equation}
	
	\section{Estimation of Shock Force and Rise Time}
	\label{app:estimation}
	
	The mechanical force acting on a single molecule during laser
	shock peening (LSP) can be estimated from the shock pressure~$P$
	and an effective molecular cross-sectional area $A\sim d^2$,
	where $d$ is a typical intermolecular distance:
	\begin{equation}
		f_{\max} = P\,d^2.
		\label{eq:force_estimate}
	\end{equation}
	
	For a small molecule with $d\sim3\,\text{\AA}$ and pressures
	characteristic of nanosecond LSP,
	$P\sim1$--$10\,\mathrm{GPa}$~\cite{Fabbro1990}, and
	femtosecond LSP, $P\sim100$--$300\,\mathrm{GPa}$,
	Eq.~\eqref{eq:force_estimate} gives the approximate range
	\begin{equation}
		f_{\max}\sim 0.09\text{--}25\,\mathrm{nN}.
	\end{equation}
	
	The shock-front rise time can be estimated from the
	shock-front thickness~$d_{\mathrm{front}}$ and the shock
	propagation velocity~$v_s$. Experimental measurements suggest
	that the shock front spans only a few molecular
	diameters~\cite{Dlott1998}, $d_{\mathrm{front}}\sim1\,\mathrm{nm}$,
	whereas typical shock velocities in condensed matter are
	$v_s\sim3$--$6\,\mathrm{km\,s^{-1}}$~\cite{Dlott1998,DlottFayer1990}.
	The time for the shock front to sweep across a molecule is
	therefore
	\begin{equation}
		\tau
		= \frac{d_{\mathrm{front}}}{v_s}
		\sim
		\frac{1\times10^{-9}\,\mathrm{m}}
		{5\times10^{3}\,\mathrm{m\,s^{-1}}}
		= 200\,\mathrm{fs}.
		\label{eq:tau_estimate}
	\end{equation}
	
	\bibliography{vibronic_refs}

@article{Beyer_chemrev,
	author  = {Beyer, Martin K. and Clausen-Schaumann, Hauke},
	title   = {Mechanochemistry: The Mechanical Activation of Covalent Bonds},
	journal = {Chemical Reviews},
	volume  = {105},
	number  = {8},
	pages   = {2921-2948},
	year    = {2005},
	month   = {07},
	issn    = {0009-2665},
	doi     = {10.1021/cr030697h},
	url     = {https://doi.org/10.1021/cr030697h},
}

@article{Pagola_review,
	AUTHOR  = {Pagola, Silvina},
	TITLE   = {Outstanding Advantages, Current Drawbacks, and Significant Recent Developments in Mechanochemistry: A Perspective View},
	JOURNAL = {Crystals},
	VOLUME  = {13},
	YEAR    = {2023},
	NUMBER  = {1},
	PAGES   = {124},
	URL     = {https://www.mdpi.com/2073-4352/13/1/124},
	ISSN    = {2073-4352},
	DOI     = {10.3390/cryst13010124}
}

@article{RibasArino2012,
	author  = {Ribas-Arino, Jordi and Marx, Dominik},
	title   = {Covalent Mechanochemistry: Theoretical Concepts and Computational Tools with Applications to Molecular Nanomechanics},
	journal = {Chemical Reviews},
	year    = {2012},
	volume  = {112},
	number  = {10},
	pages   = {5412--5487},
	doi     = {10.1021/cr200399q}
}

@article{Stauch2014,
	author  = {Stauch, Tim and Dreuw, Andreas},
	title   = {A quantitative quantum-chemical analysis tool for the distribution of mechanical force in molecules},
	journal = {The Journal of Chemical Physics},
	year    = {2014},
	volume  = {140},
	number  = {13},
	pages   = {134107},
	doi     = {10.1063/1.4870334}
}

@article{Quapp2017,
	author = {Quapp, Wolfgang and Bofill, Josep Maria and Ribas-Ari{\~n}o, Jordi},
	title = {Analysis of the Acting Forces in a Theory of Catalysis
	and Mechanochemistry},
	journal = {The Journal of Physical Chemistry A},
	volume = {121},
	number = {14},
	pages = {2820-2838},
	year = {2017},
	month = {03},
	issn = {1089-5639},
	doi = {10.1021/acs.jpca.7b00022},
	url = {https://doi.org/10.1021/acs.jpca.7b00022},
}

@article{Quapp2016,
	author  = {Quapp, Wolfgang and Bofill, Josep Maria},
	title   = {A contribution to a theory of mechanochemical pathways by means of Newton trajectories},
	journal = {Theoretical Chemistry Accounts},
	year    = {2016},
	volume  = {135},
	pages   = {113},
	doi     = {10.1007/s00214-016-1880-2}
}

@article{Sheridan2019,
	author = {Sheridan, Steven and Gr{\"a}ter, Frauke and Daday, Csaba},
	title = {How Fast Is Too Fast in Force-Probe Molecular Dynamics
	Simulations?},
	journal = {The Journal of Physical Chemistry B},
	volume = {123},
	number = {17},
	pages = {3658-3664},
	year = {2019},
	month = {04},
	issn = {1520-6106},
	doi = {10.1021/acs.jpcb.9b01251},
	url = {https://doi.org/10.1021/acs.jpcb.9b01251},
}

@article{Marcus1956,
	author  = {Marcus, R. A.},
	title   = {On the Theory of Oxidation-Reduction Reactions Involving Electron Transfer. I},
	journal = {The Journal of Chemical Physics},
	year    = {1956},
	volume  = {24},
	number  = {5},
	pages   = {966--978},
	doi     = {10.1063/1.1742723}
}

@article{cravotto2013ultrasound,
	author  = {Cravotto, Giancarlo and Gaudino, Emanuela Calcio and Cintas, Pedro},
	title   = {On the mechanochemical activation by ultrasound},
	journal = {Chemical Society Reviews},
	volume  = {42},
	number  = {18},
	pages   = {7521--7534},
	year    = {2013},
	doi     = {10.1039/C2CS35456J}
}

@article{ballmilling2025primer,
	author = {Speight, Isaiah R. and Ardila-Fierro, Karen J. and Hern{\'{a}}ndez, Jos{\'{e}} G. and Emmerling, Franziska and Michalchuk, Adam A. L. and Garc{\'i}a, Felipe and Colacino, Evelina and Mack, James},
	title = {Ball milling for mechanochemical reactions},
	journal = {Nature Reviews Methods Primers},
	volume = {5},
	pages = {29},
	year = {2025},
	doi = {10.1038/s43586-025-00401-2}
}

@article{neuman2008single,
	author  = {Neuman, Keir C. and Nagy, Attila},
	title   = {Single-molecule force spectroscopy: optical tweezers, magnetic tweezers and atomic force microscopy},
	journal = {Nature Methods},
	volume  = {5},
	number  = {6},
	pages   = {491--505},
	year    = {2008},
	doi     = {10.1038/nmeth.1218}
}

@article{rief1997reversible,
	author  = {Rief, Matthias and Gautel, Mathias and Oesterhelt, Filipp and Fernandez, Julio M. and Gaub, Hermann E.},
	title   = {Reversible Unfolding of Individual Titin Immunoglobulin Domains by AFM},
	journal = {Science},
	volume  = {276},
	number  = {5315},
	pages   = {1109--1112},
	year    = {1997},
	doi     = {10.1126/science.276.5315.1109}
}

@article{kellermayer1997folding,
	author  = {Kellermayer, Miklos S. Z. and Smith, Steven B. and Granzier, Henk L. and Bustamante, Carlos},
	title   = {Folding-Unfolding Transitions in Single Titin Molecules Characterized with Laser Tweezers},
	journal = {Science},
	volume  = {276},
	number  = {5315},
	pages   = {1112--1116},
	year    = {1997},
	doi     = {10.1126/science.276.5315.1112}
}

@article{dlott2011new,
	author  = {Dlott, Dana D.},
	title   = {New Developments in the Physical Chemistry of Shock Compression},
	journal = {Annual Review of Physical Chemistry},
	volume  = {62},
	pages   = {575--597},
	year    = {2011},
	doi     = {10.1146/annurev.physchem.012809.103514}
}

@article{fairand1979laser,
	author  = {Fairand, B. P. and Clauer, A. H.},
	title   = {Laser generation of high-amplitude stress waves in materials},
	journal = {Journal of Applied Physics},
	volume  = {50},
	pages   = {1497--1502},
	year    = {1979}
}

@article{peyre1995laser,
	author  = {Peyre, P. and Fabbro, R.},
	title   = {Laser shock processing: a review of the physics and applications},
	journal = {Optical and Quantum Electronics},
	volume  = {27},
	number  = {12},
	pages   = {1213--1229},
	year    = {1995}
}

@article{alonso2024wandering,
	author  = {Alonso, Mercedes and Bettens, Tom and Eeckhoudt, Jochen and Geerlings, Paul and De Proft, Frank},
	title   = {Wandering through quantum-mechanochemistry: from concepts to reactivity and switches},
	journal = {Physical Chemistry Chemical Physics},
	volume  = {26},
	pages   = {21-{}-35},
	year    = {2024},
	doi     = {10.1039/D3CP04907H}
}

@article{stauch2016advances,
	author  = {Stauch, Tim and Dreuw, Andreas},
	title   = {Advances in Quantum Mechanochemistry: Electronic Structure Methods and Force Analysis},
	journal = {Chemical Reviews},
	volume  = {116},
	number  = {22},
	pages   = {14137-{}-14180},
	year    = {2016},
	doi     = {10.1021/acs.chemrev.6b00458}
}

@article{Konda2011,
	author  = {Konda, S. S. and Brantley, J. N. and Bielawski, C. W. and Makarov, D. E.},
	title   = {Chemical reactions modulated by mechanical stress: extended {Bell} theory},
	journal = {The Journal of Chemical Physics},
	year    = {2011},
	volume  = {135},
	number  = {16},
	pages   = {164103},
	doi     = {10.1063/1.3656367}
}

@article{RibasArino2009,
	author  = {Ribas-Arino, Jordi and Shiga, Motoyuki and Marx, Dominik},
	title   = {Understanding Covalent Mechanochemistry},
	journal = {Angewandte Chemie International Edition},
	year    = {2009},
	volume  = {48},
	number  = {23},
	pages   = {4190--4193},
	doi     = {10.1002/anie.200900673}
}

@article{Beyer2000,
	author  = {Beyer, Martin K.},
	title   = {The mechanical strength of a covalent bond calculated by density functional theory},
	journal = {The Journal of Chemical Physics},
	year    = {2000},
	volume  = {112},
	number  = {17},
	pages   = {7307--7312},
	doi     = {10.1063/1.481330}
}

@article{Wolinski2009,
	author     = {Wolinski, Krzysztof and Baker, Jon},
	title      = {Theoretical predictions of enforced structural changes in molecules},
	journal    = {Molecular Physics},
	volume     = {107},
	number     = {22},
	pages      = {2403--2417},
	year       = {2009},
	publisher = {Taylor \& Francis},
	doi       = {10.1080/00268970903321348},
	url       = {https://doi.org}
}

@article{Weiss2024,
	author  = {Wei{\ss}, Rahel and Zeller, Felix and Neudecker, Tim},
	title   = {Calculating high-pressure vibrational frequencies analytically with the extended hydrostatic compression force field approach},
	journal = {The Journal of Chemical Physics},
	year    = {2024},
	volume  = {160},
	pages   = {084101},
	doi     = {10.1063/5.0189887}
}

@article{Stauch2020,
	author  = {Stauch, Tim},
	title   = {A mechanochemical model for the simulation of molecules and molecular crystals under hydrostatic pressure},
	journal = {The Journal of Chemical Physics},
	year    = {2020},
	volume  = {153},
	pages   = {134503},
	doi     = {10.1063/5.0022580}
}

@article{Smalo2013,
	author  = {Smal{\o}, Hans Sverre and Uggerud, Einar},
	title   = {Breaking covalent bonds using mechanical force, which bond breaks?},
	journal = {Molecular Physics},
	year    = {2013},
	volume  = {111},
	number  = {9-11},
	pages   = {1563--1573},
	doi     = {10.1080/00268976.2013.811554}
}

@article{Pill2019,
	author  = {Pill, Michael F. and East, Allan L. L. and Marx, Dominik and Beyer, Martin K. and Clausen-Schaumann, Hauke},
	title   = {Mechanical Activation Drastically Accelerates Amide Bond Hydrolysis, Matching Enzyme Activity},
	journal = {Angewandte Chemie International Edition},
	year    = {2019},
	volume  = {58},
	number  = {29},
	pages   = {9787--9790},
	doi     = {10.1002/anie.201902752}
}

@article{Ong2019,
	author = {Ong, Mitchell T. and Leiding, Jeff and Tao, Hongli and Virshup, Aaron M. and Mart{\'{i}}nez, Todd J.},
	title = {First Principles Dynamics and Minimum Energy Pathways for Mechanochemical Ring Opening of Cyclobutene},
	journal = {Journal of the American Chemical Society},
	volume = {131},
	number = {18},
	pages = {6377-6379},
	year = {2009},
	month = {04},
	doi = {10.1021/ja8095834},
	url = {https://doi.org/10.1021/ja8095834},
}

@article{Stauch2017,
	author = {Stauch, Tim and Dreuw, Andreas},
	title = {Force-induced retro-click reaction of triazoles competes with adjacent single-bond rupture},
	journal = {Chemical Science},
	volume = {8},
	number = {8},
	pages = {5567-5575},
	year = {2017},
	month = {08},
	issn = {2041-6520},
	doi = {10.1039/c7sc01562c},
	url = {https://doi.org/10.1039/c7sc01562c},
}

@article{Nepal2021,
	author = {Kedziora, Gary S. and Moller, James and Berry, Rajiv and Nepal, Dhriti},
	title = {Ab initio molecular dynamics modeling of single polyethylene chains: Scission kinetics and influence of radical under mechanical strain},
	journal = {The Journal of Chemical Physics},
	volume = {155},
	number = {2},
	pages = {024102},
	year = {2021},
	month = {07},
	issn = {0021-9606},
	doi = {10.1063/5.0047371},
	url = {https://doi.org/10.1063/5.0047371},
}

@article{Dopieralski2013,
	author  = {Dopieralski, Przemyslaw and Ribas-Arino, Jordi and Anjukandi, Padmesh and Krupicka, Martin and Kiss, Jozsef and Marx, Dominik},
	title   = {The Janus-faced role of external forces in mechanochemical disulfide bond cleavage},
	journal = {Nature Chemistry},
	year    = {2013},
	volume  = {5},
	pages   = {685--691},
	doi     = {10.1038/nchem.1676}
}

@article{Smalo2014,
	author  = {Smal{\o}, Hans Sverre and Rybkin, Vladimir V. and Klopper, Wim and Helgaker, Trygve and Uggerud, Einar},
	title   = {Mechanochemistry: The Effect of Dynamics},
	journal = {The Journal of Physical Chemistry A},
	year    = {2014},
	volume  = {118},
	number  = {36},
	pages   = {7683--7694},
	doi     = {10.1021/jp504959z}
}

@article{condon1926theory,
	title={A theory of intensity distribution in band systems},
	author={Condon, Edward},
	journal={Physical Review},
	volume={28},
	number={6},
	pages={1182},
	year={1926},
	publisher={APS}
}

@article{frank1975electron,
	title   = {Electron-vibrational interactions in polyatomic molecules},
	author  = {Frank-Kamenetskii, MD and Lukashin, Aleksey Viktorovich},
	journal = {Soviet Physics Uspekhi},
	volume  = {18},
	number  = {6},
	pages   = {391--409},
	year    = {1975}
}

@article{Huang1950TheoryOL,
	title={Theory of light absorption and non-radiative transitions in F-centres},
	author={Kun Huang and A. Farrer Rhys},
	journal={Proceedings of the Royal Society of London. Series A. Mathematical and Physical Sciences},
	year={1950},
	volume={204},
	pages={406 - 423},
	url={https://api.semanticscholar.org/CorpusID:97930503}
}

@article{Rybkin2017,
	author    = {Rybkin, Vladimir V.},
	title     = {Franck-Condon Theory of Quantum Mechanochemistry},
	journal   = {The Journal of Physical Chemistry A},
	volume    = {121},
	number    = {30},
	pages     = {5758--5762},
	year      = {2017},
	doi       = {10.1021/acs.jpca.7b05264},
	publisher = {American Chemical Society},
}

@article{Fabbro1990,
	author    = {Fabbro, R. and Fournier, J. and Ballard, P.
	and Devaux, D. and Virmont, J.},
	title     = {Physical study of laser-produced plasma in confined geometry},
	journal   = {Journal of Applied Physics},
	volume    = {68},
	number    = {2},
	pages     = {775--784},
	year      = {1990},
	doi       = {10.1063/1.346783},
}

@article{Dlott1998,
	author    = {Dlott, Dana D.},
	title     = {The New Wave in Shock Waves},
	journal   = {The Journal of Physical Chemistry B},
	volume    = {102},
	number    = {12},
	pages     = {2121--2130},
	year      = {1998},
	doi       = {10.1021/jp973404v},
}

@article{DlottFayer1990,
	author    = {Dlott, D. D. and Fayer, M. D.},
	title     = {Shocked molecular solids: vibrational up pumping,
	defect hot spot formation, and the onset of chemistry},
	journal   = {The Journal of Chemical Physics},
	volume    = {92},
	pages     = {3798--3812},
	year      = {1990},
	doi       = {10.1063/1.457838},
}

@ARTICLE{1981UsFiN.134..185R,
	author = {{Rusanov}, V.~D. and {Fridman}, A.~A. and {Sholin}, G.~V.},
	title = "{The physics of a chemically active plasma with nonequilibrium vibrational excitation of molecules}",
	journal = {Uspekhi Fizicheskikh Nauk},
	year = 1981,
	month = jun,
	volume = {134},
	pages = {185-235},
	adsurl = {https://ui.adsabs.harvard.edu/abs/1981UsFiN.134..185R}
}

@article{macheret1993chemphys,
	author  = {Macheret, S. O. and Rich, J. W.},
	title   = {Nonequilibrium dissociation rates behind strong shock waves:
	classical model},
	journal = {Chemical Physics},
	volume  = {174},
	number  = {1},
	pages   = {25--43},
	year    = {1993},
	doi     = {10.1016/0301-0104(93)80049-F}
}

@article{bell1978models,
	title   = {Models for the specific adhesion of cells to cells},
	author  = {Bell, George I.},
	journal = {Science},
	volume  = {200},
	number  = {4342},
	pages   = {618--627},
	year    = {1978},
	doi     = {10.1126/science.347575}
}

@article{tokmakoff1993chemical,
	author  = {Tokmakoff, Andrei and Fayer, M. D. and Dlott, Dana D.},
	title   = {Chemical Reaction Initiation and Hot Spot Formation in Shocked Energetic Molecular Materials},
	journal = {The Journal of Physical Chemistry},
	volume  = {97},
	number  = {9},
	pages   = {1901--1913},
	year    = {1993},
	doi     = {10.1021/j100111a039}
}

@book{cohen-tannoudji_qm_vol1,
	author    = {Claude Cohen-Tannoudji and Bernard Diu and Franck Lalo{\"e}},
	title     = {Quantum Mechanics, Volume 1: Basic Concepts, Tools, and Applications},
	publisher = {Wiley-VCH},
	year      = {1977},
	edition   = {1st},
	isbn      = {978-0471164333}
}

@book{lapierre2018getting,
	title={Getting started in quantum optics},
	author={LaPierre, Ray},
	year={2018},
	publisher={Springer}
}

@article{Eickmann2026,
	author    = {Eickmann, Jan-Lucas and Luo, Kai-Hong and Roiz, Mikhail and Lammers, Jonas and Atzeni, Simone and Pandey, Cheeranjiv and L{\"u}tkewitte, Florian and Shirazi, Reza G. and Schlue, Fabian and Brecht, Benjamin and Rybkin, Vladimir V. and Stefszky, Michael and Silberhorn, Christine},
	title     = {Bridging chemistry and Gaussian boson sampling: a photonic hierarchy of approximations for molecular vibronic spectra},
	journal   = {npj Quantum Information},
	year      = {2026},
	volume    = {12},
	number    = {1},
	pages     = {89},
	doi       = {10.1038/s41534-026-01250-x},
	issn      = {2056-6387}
}

@article{Evans1996,
	author  = {Evans, R. and Badger, A. D. and Fallies, F. and Mahdieh, M. and Hall, T. A. and Audebert, P. and Geindre, J. P. and Gauthier, J. C. and Mysyrowicz, A. and Grillon, G. and Antonetti, A.},
	title   = {Time- and Space-Resolved Optical Probing of Femtosecond-Laser-Driven Shock Waves in Aluminum},
	journal = {Physical Review Letters},
	year    = {1996},
	volume  = {77},
	number  = {16},
	pages   = {3359--3362},
	doi     = {10.1103/PhysRevLett.77.3359}
}

@article{James2012,
	author    = {James, Stuart L. and Adams, Christopher J. and Bolm, Carsten and Braga, Dario and Collier, Paul and Fri{\v{s}}{\v{c}}i{\'c}, Tomislav and Grepioni, Fabrizia and Harris, Kenneth D. M. and Hyett, Geoff and Jones, William and Krebs, Antje and Mack, Jamie and Maini, Lusi and Orpen, A. Guy and Parkin, Ivan P. and Shearouse, William C. and Steed, Jonathan W. and Waddell, Krishna C.},
	title     = {Mechanochemistry},
	journal   = {Chemical Society Reviews},
	year      = {2012},
	volume    = {41},
	number    = {1},
	pages     = {413--447},
	doi       = {10.1039/C1CS15171A},
	publisher = {Royal Society of Chemistry}
}

@book{Wilson1955,
	author    = {Wilson, E. Bright and Decius, J. C. and Cross, Paul C.},
	title     = {Molecular Vibrations: The Theory of Infrared and Raman Vibrational Spectra},
	publisher = {McGraw--Hill},
	address   = {New York},
	year      = {1955},
	pages     = {388}
}

@article{Sun2018,
	author  = {Sun, Qiming and Berkelbach, Timothy C. and Blunt, Nick S. and Booth, George H. and Guo, Sheng and Li, Zack and Liu, Junzi and McClain, James D. and Sayfutyarova, Elmira R. and Sharma, Sandeep and Wouters, Sebastian and Chan, Garnet Kin-Lic},
	title   = {{PySCF}: The Python-Based Simulations of Chemistry Framework},
	journal = {WIREs Computational Molecular Science},
	year    = {2018},
	volume  = {8},
	number  = {1},
	pages   = {e1340},
	doi     = {10.1002/wcms.1340}
}

@article{Becke1993,
	author  = {Becke, Axel D.},
	title   = {Density-Functional Thermochemistry. III. The Role of Exact Exchange},
	journal = {The Journal of Chemical Physics},
	year    = {1993},
	volume  = {98},
	number  = {7},
	pages   = {5648--5652},
	doi     = {10.1063/1.464913}
}

@article{Lee1988,
	author  = {Lee, Chengteh and Yang, Weitao and Parr, Robert G.},
	title   = {Development of the Colle--Salvetti Correlation-Energy Formula into a Functional of the Electron Density},
	journal = {Physical Review B},
	year    = {1988},
	volume  = {37},
	number  = {2},
	pages   = {785--789},
	doi     = {10.1103/PhysRevB.37.785}
}

@article{Krishnan1980,
	author  = {Krishnan, R. and Binkley, J. S. and Seeger, R. and Pople, J. A.},
	title   = {Self-Consistent Molecular Orbital Methods. XX. A Basis Set for Correlated Wave Functions},
	journal = {The Journal of Chemical Physics},
	year    = {1980},
	volume  = {72},
	number  = {1},
	pages   = {650--654},
	doi     = {10.1063/1.438955}
}
	
\end{document}